\documentclass[twocolumn,aps,showpacs,showkeys,tightenlines,superscriptaddress]{revtex4}
\usepackage{epsfig,bm}
\usepackage[dvips]{color}
\newcommand{\vi}[1]{\mbox{\boldmath $#1$}}
\newcommand{\vis}[1]{\mbox{\boldmath ${\scriptstyle #1}$}}
\begin{document}

\title{Elastic and total reaction cross
sections of oxygen isotopes in Glauber theory}

\author{B. Abu-Ibrahim}
\affiliation{RIKEN Nishina Center, RIKEN, Wako-shi, 
Saitama 351-0198, Japan}
\affiliation{Department of Physics, Cairo University, 
Giza 12613, Egypt}
\author{S. Iwasaki\footnote{Present address: Intertrade Co., Ltd.,
Chuo-ku, Tokyo 104-0032, Japan}}
\affiliation{Graduate School of Science and Technology, 
Niigata University, Niigata 950-2181, Japan}
\author{W. Horiuchi}
\affiliation{Graduate School of Science and Technology, 
Niigata University, Niigata 950-2181, Japan}
\author{A. Kohama}
\affiliation{RIKEN Nishina Center, RIKEN, Wako-shi, 
Saitama 351-0198, Japan}
\author{Y. Suzuki}
\affiliation{Department of Physics, 
Graduate School of Science and Technology, 
Niigata University, Niigata 950-2181, Japan}


\begin{abstract}
We systematically calculate the total reaction cross sections of 
oxygen isotopes, 
$^{15-24}$O, on a $^{12}$C target at high energies using the 
Glauber theory. 
The oxygen isotopes are described with Slater determinants 
generated from a phenomenological mean-field potential. The 
agreement between theory 
and experiment is generally good, but a sharp increase 
of the reaction cross sections 
from $^{21}$O to $^{23}$O remains unresolved. 
To examine the sensitivity of the diffraction pattern of 
elastic scattering to the nuclear surface, 
we study the differential elastic-scattering cross sections of 
proton\,-$^{20,21,23}$O at the incident energy of 300\,MeV 
by calculating the full Glauber amplitude. 
\end{abstract}

\maketitle

\section{Introduction}

Studies on neutron-rich unstable nuclei have been attracting much 
attention both experimentally and theoretically. 
These studies are motivated, for example, by that 
we want to understand the nuclear structure and excitation mode 
of the neutron-rich nuclei as well as a role played by them 
in forming heavy elements in stars.   
Binding energy, radius and density distribution, among others, 
are basic quantities to determine the nuclear property. 
Reactions of unstable neutron-rich 
nuclei with a proton target are, therefore, of current 
interest as they are at present a major means to probe the 
matter densities of exotic nuclei, particularly the region 
of the nuclear surface.   
If one appropriately selects the incident energies, protons could 
be more sensitive to the neutron distributions than to the 
proton distributions of nuclei.

The Glauber theory~\cite{Glauber} offers 
a powerful and handy framework 
for the description of high energy nuclear reactions. This theory  
describes proton-nucleus reactions very well. Usually a 
calculation based on the Glauber theory 
is performed using a one-body  
density calculated from a nuclear wave function. 
However, Bassel {\it et al.}~\cite{bassel}
calculated Glauber's scattering amplitude 
using a Slater determinant wave function, 
and were able to examine the effect of Pauli blocking, 
which is impossible to discuss using just the 
one-body density.

Recently, we have analyzed the total reaction cross sections of 
carbon isotopes on both $^{12}$C~\cite{hsbk} and proton~\cite{bhks} 
targets using the Glauber theory. 
The densities of the carbon isotopes are constructed from 
Slater determinants generated from a phenomenological mean-field 
potential or two types of dynamical models, core+$n$ and core+$n$+$n$
models, which go beyond the mean-field model.

Data on the cross sections 
of the oxygen isotopes up to the dripline nucleus $^{24}$O 
are available at high energies~\cite{ozawa2001,ozawa2001a}: 
The cross sections show a gentle increase with increasing neutron 
number from $^{16}$O to $^{21}$O and then an abrupt 
increase up to $^{23}$O. 
On the other hand, several theoretical studies on the matter radii 
predict a mild increase of the radii with increasing neutron number. 
The calculations have been performed for the isotopes 
up to $N = 12$ \cite{masui}, and much further 
\cite{lalazi,enyo,saviankou,nakada}. 
As the spin-parity systematics of 
the ground states of the oxygen isotopes indicates that 
the $d_{5/2}$ single-particle orbit seems fairly stable,  
we first analyze the cross section data 
using a model similar to that of the previous study, that is, 
assuming the mean-field potentials that reproduce the 
nucleon separation energies. 
The interest of this 
analysis is to examine the extent to which the theory can explain 
the characteristic behavior of the cross sections. 

One of the advantages of the Glauber theory is that the same 
input as used in the reaction cross section calculation is 
readily applicable for the 
calculation of the differential cross section of elastic-scattering. 
The diffraction pattern in the differential 
cross section is expected to depend on the diffuseness of the 
nuclear surface. 
Therefore we also study elastic scattering of the 
oxygen isotopes on a 
proton target. The case of $p$\,-$^{16}$O scattering
serves as a testing ground of our model. We then predict the
differential cross sections of $p$\,-$^{20,21,23}$O elastic 
scattering. 
The differential elastic-scattering cross section data 
of $p$-$^{20}$O is available only at low energy, 
such as $\sim30$ MeV per nucleon~\cite{jewell}.

This paper is organized as follows. 
The reaction models for the calculation 
of the total reaction and elastic-scattering 
cross sections are recapitulated in Sect.~\ref{psf}. The 
construction of the densities of the oxygen isotopes is 
explained in Sect.~\ref{density}. 
We present the calculation of the total reaction cross sections of 
the oxygen isotopes in Sect.~\ref{reac}. 
The differential cross sections for neutron-rich oxygen isotopes 
are given in Sect.~\ref{elastic}. 
A summary is given in Sect.~\ref{summary}. 
Appendix~A presents the densities of the oxygen isotopes 
which are obtained using the 
Slater determinants constructed from the harmonic-oscillator 
single-particle orbits. 
Appendix~B explains the calculation of the $p$\,-nucleus 
elastic-scattering amplitude using a Slater determinant.

\section{Phase shift functions in the Glauber model}   
\label{psf}

The Glauber theory provides us with an excellent framework to 
describe high energy reactions~\cite{Glauber}. 
The $p$\,-nucleus elastic-scattering amplitude is given by
\begin{equation}
F({\vi q})=\frac{iK}{2 \pi}\int 
d{\vi b}\,{\rm e}^{i{\vis {q\cdot b}}}
\left(1-{\rm e}^{i\chi_N({\vis b})}\right),
\label{glamp}
\end{equation}
where ${\vi b}$ is the impact parameter vector
perpendicular to the beam ($z$) direction, 
${\vi K}$ the initial momentum of the relative motion,  
and ${\vi q}$ the momentum 
transferred from the projectile to the target.
Here we neglect the $p$\,-nucleus Coulomb potential, 
because its effect appears 
only in the extreme forward direction for a light target. 
The nuclear part of the optical phase-shift function 
is given by 
\begin{eqnarray}
   {\rm e}^{i\chi_N({\vis b})}
   &\!\!=\!\!& \langle\Psi_{0}\vert
     \prod_{i=1}^{A}
     \Big[1 - \frac{1+\tau_{3_i}}{2}
              \Gamma_{pn}({\vi b} - {\vi s}_{i})
\nonumber \\
     & & \ \ \qquad \qquad 
          -   \frac{1-\tau_{3_i}}{2}
              \Gamma_{pp}({\vi b} - {\vi s}_{i}) 
     \Big]|\Psi_{0} \rangle, 
\label{opsf}
\end{eqnarray}
where $\Psi_{0}$ is the intrinsic (translation-invariant) 
$A$-nucleon wave function of the nuclear ground state, 
and ${\vi s}_{i}$ is the projection onto the $xy$-plane 
of the nucleon coordinate relative to 
the center-of-mass of the nucleus. 
Here $\tau_{3_i}$ is $1$ for neutron and $-1$ for proton.

The profile function, $\Gamma_{pN}$, 
for the $pp$ and $pn$ elastic scatterings, is usually parameterized 
in the form
\begin{equation}
  \Gamma_{pN}({\vi b}) =
  \frac{1-i\alpha_{pN}}{4\pi\beta_{pN}}\,\,
  \sigma_{pN}^{\rm tot}\, {\rm exp}\left(-\frac{{\vi b}^2} {2\beta_{pN}}\right),
\label{gfn}
\end{equation}
where $\alpha_{pN}$ is the ratio of 
the real to the imaginary part of the 
$pN$ scattering amplitude in the forward direction, 
$\sigma_{pN}^{\rm tot}$ is the $pN$ total cross sections, 
and $\beta_{pN}$ is the slope parameter of the 
$pN$ differential  cross section. 
These parameters for different incident energies 
are tabulated in Refs.~\cite{hsbk,bhks}.

The differential  cross section of the 
$p$-nucleus elastic-scattering is given by
\begin{equation}
\frac{d\sigma}{d\Omega}=\big|F({\vi q})\big|^{2},
\end{equation}
and the total reaction cross section of the $p$\,-nucleus 
collision is calculated from 
\begin{equation}
  \sigma_{\rm R} = \int{d{\vi b}\,
  \left(1-\big|{\rm e}^{i\chi_N({\vis b})}\big|^{2}\right)}.
\label{reaccs}
\end{equation}

In the optical limit approximation (OLA), 
the $p$\,-nucleus scattering phase-shift function  
is simply given using the proton density $\rho_{p}({\vi r})$ 
and the neutron density $\rho_{n}({\vi r})$ as follows 
\begin{equation}
   {\rm e}^{i\chi_{\rm OLA}({\vis b})}
   = \exp \left[i\chi_{p}({\vi b})+
                i\chi_{n}({\vi b}) \right],
\label{ola}
\end{equation}
where $\chi_{p}$ ($\chi_{n}$) denotes 
the phase shift due to the protons (neutrons) inside the nucleus 
\begin{eqnarray}
i\chi_{p}({\vi b}) &=& -{\int{d{\vi r} }} \rho_{p}({\vi r})
                        \Gamma_{pp}({\vi b} - {\vi s}), 
\nonumber\\
i\chi_{n}({\vi b}) &=& -{\int{d{\vi r} }} \rho_{n}({\vi r})
                        \Gamma_{pn}({\vi b} - {\vi s}). 
\label{chi}
\end{eqnarray}

We will evaluate Eq.~(\ref{opsf}) completely for a $\Psi_0$ that 
is given by a Slater determinant comprising harmonic-oscillator 
single-particle orbits, and use it in Sect.~\ref{elastic}.

\section{Densities of Oxygen isotopes}
\label{density}

\begin{figure*}[t]
\epsfig{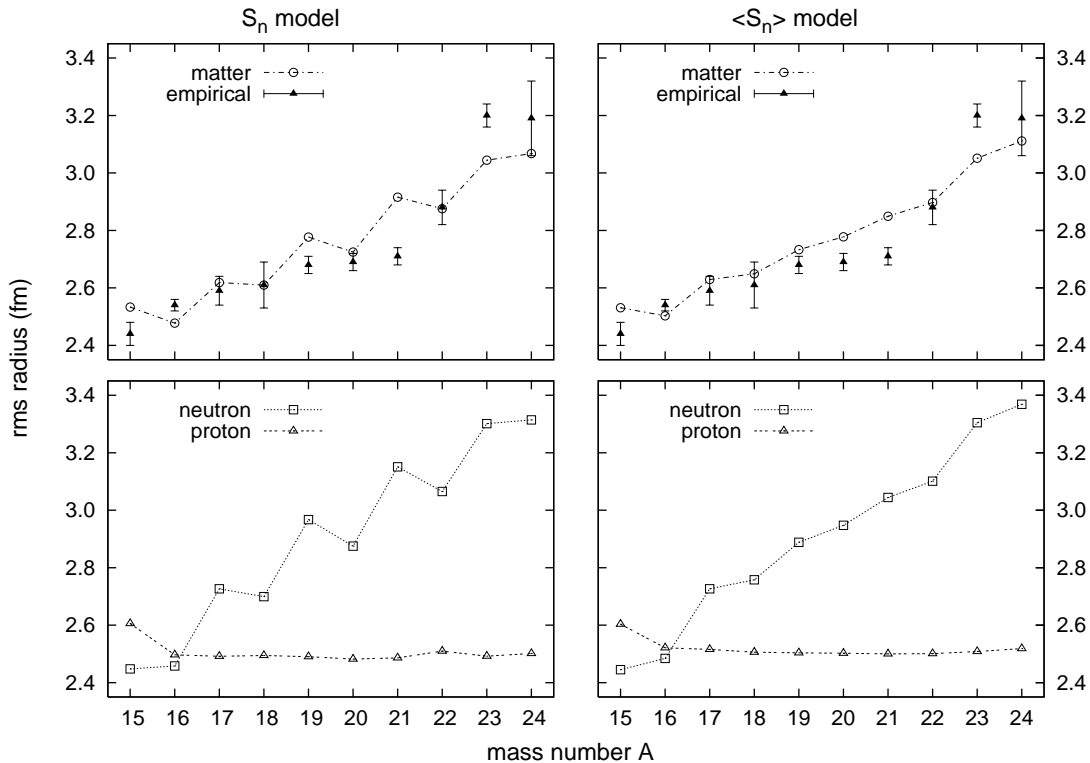}
\caption{The matter, neutron and 
proton root-mean-square radii of the oxygen isotopes 
calculated with the $S_n$ and $\langle S_n \rangle$ models. 
The empirical values are taken from Ref.~\cite{ozawa2001}.}
\label{rms.size}
\end{figure*}

Compared to the carbon isotopes, where the competition 
of the 1$s_{1/2}$ and 
0$d_{5/2}$ orbits appears to play an important role 
to determine their 
structure, the spin-parity systematics of the oxygen isotopes 
indicates
that the 0$d_{5/2}$ orbit is lower than the 1$s_{1/2}$ orbit 
in the ground states up to $^{22}$O. The configurations for the 
oxygen isotopes are thus assumed to be given according to the 
shell-model as follows: The nucleus $^{16}$O is a 
doubly magic nucleus occupying the $0s_{1/2}$, $0p_{3/2}$ and 
$0p_{1/2}$ orbits, and 
$^{15}$O has a neutron hole in the $0p_{1/2}$ orbit. The nucleus 
$^A$O with 16$<\!A\!\leq$\,22 has $A\!-\!16$ valence 
neutrons in the $0d_{5/2}$ orbit,  
and for $A$=23, 24 it has six neutrons in the $0d_{5/2}$ orbit 
as well as $A\!-\!22$ neutrons 
in the $1s_{1/2}$ orbit. We assume the spin-parity of 
the ground state of $^{21}$O is ${\frac{5}{2}}^+$ though 
it is not yet confirmed experimentally. 

The single-particle 
orbits are generated from the mean-field potential containing 
central and spin-orbit potentials. The Coulomb potential 
is added for protons. 
The strength of the spin-orbit potential is set to follow the 
standard value, whereas the strength of the central part 
is chosen so as to reproduce the separation energy of the last 
nucleon. See Ref.~\cite{hsbk} for details. 
This prescription is called an ``$S_n$ model'' hereafter. 
This model apparently ignores the pairing effect, which gives 
larger separation energy for the even-$N$ nucleus than 
for the odd-$N$ nucleus. Therefore the $S_n$ model tends to 
predict a too large size for the odd-$N$ isotope. To remedy this 
problem, we also test another one, called an 
``$\langle S_n \rangle$ model'', which fits the 
average separation energy 
for the nucleons in the last orbit. For example, in the case 
of $^{19}$O in which three neutrons occupy the $0d_{5/2}$ orbit, 
the average neutron separation energy to be fitted 
is one third of a sum of the neutron separation 
energies of $^{17,18,19}$O, and likewise the average proton 
separation energy to be fitted is one half of a sum of the 
proton separation energies of $^{19}$O and $^{18}$N.  
The center-of-mass motion is taken into account in order to obtain 
the intrinsic density. See Ref.~\cite{hsbk} for detail.

We also use harmonic-oscillator single-particle orbits in
Sect.~\ref{elastic} to examine the differential cross section.

Figure~\ref{rms.size} compares the matter, 
neutron and proton root-mean-square (rms) 
radii of the oxygen isotopes 
($r_m$, $r_n$, and $r_p$, respectively) calculated in  
the $S_n$ and $\langle S_n \rangle$ models. Clearly 
the strong even-odd staggering seen in the neutron radius of  
the $S_n$ model 
becomes mild in the $\langle S_n \rangle$ model, and the 
matter radii in the latter model seem to be in better 
agreement with those extracted from a model-dependent 
analysis of the interaction cross section data~\cite{ozawa2001}. 
Noticeable discrepancies appear at $^{21}$O and $^{23}$O. The theory 
predicts a smaller radius for $^{21}$O but a larger radius 
for $^{23}$O than the empirical values. We will see in 
Sect.~\ref{reac} that this discrepancy  
directly appears in the comparison of the reaction cross sections. 
The charge radii of $^{16,17,18}$O calculated 
in the $\langle S_n \rangle$ model using the 
finite size correction of the nucleon 
are found to be 2.64, 2.63, 2.62\,fm, 
which are slightly smaller than the measured values, 
2.718(21), 2.662(26), 2.727(20)\,fm~\cite{devries}, respectively.

\section{Results for total reaction cross sections}
\label{reac}

\begin{figure}[b]
\epsfig{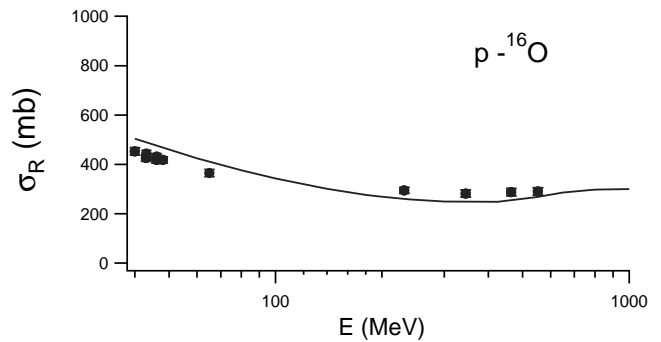}
\caption{Comparison of the $p$+$^{16}$O total 
reaction cross sections calculated in the OLA with experiment. 
The data are taken from Ref.~\cite{carlson,ingema}.}
\label{p-16O.reac}
\end{figure}

\begin{figure}[t]
\epsfig{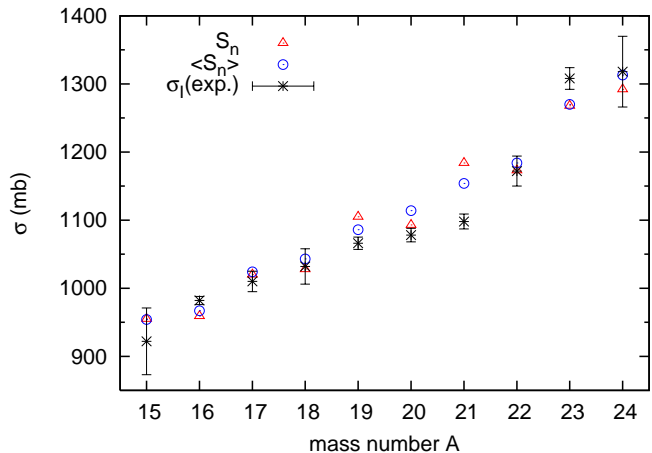}
\caption{(Color online) 
Total reaction cross sections for the oxygen isotopes incident on 
a $^{12}$C target. The incident energy is about 1\,GeV per 
nucleon except for $^{15}$O. The calculation is based on the 
NTG approximation. The experimental data are taken 
from Ref.~\cite{ozawa2001}. }
\label{reac.oxygen}
\end{figure}

Before showing results of calculation for 
the total reaction cross sections of the oxygen isotopes, 
we first compare in Fig.~\ref{p-16O.reac} the theoretical 
$p$\,-$^{16}$O total 
reaction cross section to experiment. The OLA 
approximation is used. 
The cross sections calculated at higher energies are in reasonable 
agreement with the data. A slight underestimation is consistent with 
the fact that the density of $^{16}$O gives 
a slightly smaller radius compared to the experimental value. 
The theory predicts about 15\% larger cross sections 
at lower energies, which is very similar to the one 
found in the analysis of the $p$\,-$^{12}$C total reaction cross 
section~\cite{bhks}. The reason for this discrepancy is due to the 
limitation of applying the Glauber theory to the reaction 
at the energy lower than 100\,MeV~\cite{slyv}.

Figure~\ref{reac.oxygen} displays the total 
reaction cross sections of the oxygen isotopes incident 
on a $^{12}$C target.  
The calculated values are based on the 
NTG approximation~\cite{hsbk}, which includes higher-order 
corrections missing in the OLA calculation, and 
they are compared to the interaction cross sections 
measured at incident 
energies around 1\,GeV per nucleon~\cite{ozawa2001}. The energy per 
nucleon is, however, about 710\,MeV in the case of $^{15}$O. 
The density used for $^{12}$C is the same as in Ref.~\cite{hsbk}. 
The $\langle S_n \rangle$ model seems to give slightly 
better results than the $S_n$ model. 

For a comparison between 
theory and experiment, it is important to realize 
that it is the total reaction cross section that is calculated 
and not the interaction cross section. 
The interaction cross section does not include the 
contribution from the inelastic excitation of a projectile nucleus 
to a particle-bound state, whereas the total reaction 
cross section includes this contribution. 
It is in general not trivial 
to estimate the difference between the two cross sections 
\cite{kio08}, but 
they are not expected to be very different at high energies 
\cite{ogawa,Ozawa02}.  
As already noted in the comparison of the matter radii, the 
theory can reproduce the increase of the cross 
sections up to $^{20}$O but fails to reproduce the sharp rise of the 
cross section from $^{21}$O to $^{23}$O. 
It was noted in Ref.~\cite{kanungo1} that the latter discrepancy 
cannot be explained in any theoretical models. See Fig.\,4 of 
Ref.~\cite{kanungo1} as well as the calculation 
of Ref.~\cite{brown}. 

The sharp rise in the interaction cross section at $^{23}$O 
has led to some controversial issues for the understanding of 
its origin. 
It was conjectured in Refs.~\cite{kanungo1,kanungo2} that 
a core nucleus $^{22}$O could be substantially modified 
in $^{23}$O if 
described with a $^{22}$O+$n$ model and also that 
the spin-parity of the ground state of $^{23}$O might 
be $\frac{5}{2}^+$ 
rather than $\frac{1}{2}^+$. Motivated by this, several experiments 
have been performed, which all support 
$\frac{1}{2}^+$~\cite{sauvan,cortina-gil,nociforo,elekes,schiller}, 
as adopted in the present study. 
Namely, the configuration of the ground state of $^{23}$O is 
dominated by 
a $1s_{1/2}$ neutron coupled to $^{22}$O(0$^+$) and an excited 
configuration of $(1s_{1/2})^2(0d_{5/2})^{-1}$ appears 
as a resonance 
lying very close to the $n$+$^{22}$O 
threshold~\cite{nociforo,schiller}. 
The enhancement of the cross section from $^{21}$O to $^{23}$O 
thus remains an open question to be explained. 

As another 
possibility for getting information on nuclear size, 
particularly near 
its surface, we will examine in the next section the  
$p$\,-nucleus elastic scattering.

\section{Elastic-scattering cross sections}
\label{elastic}

In the previous section we calculated the total reaction cross 
section for the oxygen isotopes using the phenomenological 
mean-field potential. 
Though the $\langle S_n \rangle$ model was found to give 
nuclear sizes that reproduce the cross sections reasonably well, 
the increase of the cross section from $^{21}$O to $^{23}$O 
was not explained. The $p$\,-$^{20}$O elastic scattering 
has recently been measured at the National Institute of 
Radiological Sciences (NIRS) in Japan~\cite{tera}. 
In this section we study the differential   
cross section of the $p$-oxygen isotope elastic-scattering in order 
to examine how sensitive the differential cross section is to 
the nuclear size. 

To this end, it is more convenient if we can adjust the nuclear size 
more flexibly. Thus we use harmonic-oscillator single-particle 
wave functions instead of the orbits determined from the separation 
energies as was done in Sect.~\ref{density}. 
The size parameters $\nu_p$ and $\nu_n$ 
of the harmonic-oscillator wave functions are first determined so as 
to reproduce the proton and neutron rms radii of the 
$\langle S_n \rangle$ model and then, by changing 
the $\nu_n$ value to fit the total reaction cross section, 
we examine 
the extent to which the differential cross section is altered. 
Table~\ref{radii} lists the radii and the total reaction 
cross sections of the oxygen isotopes for several sets of 
$\nu_p$ and $\nu_n$ values adopted in this section. 
The $r_m,\, r_n$, and $r_p$ values 
in each row containing a parenthesis correspond to the values 
of the $\langle S_n \rangle$ model.  
The density formula obtained from the harmonic-oscillator functions 
are summarized in Appendix~A.   

The calculation of the Glauber amplitude (\ref{glamp}) 
for the $p$\,-nucleus elastic scattering 
is explained in Appendix~B. 

\subsection{Test of the Glauber amplitude: $^{16}$O}
We present numerical results 
of the differential cross section 
for the $p$\,-$^{16}$O  elastic-scattering, 
and compare it to experiment to study how precisely 
we can discuss the nuclear size within the Glauber theory.

\begin{figure}[t]
\epsfig{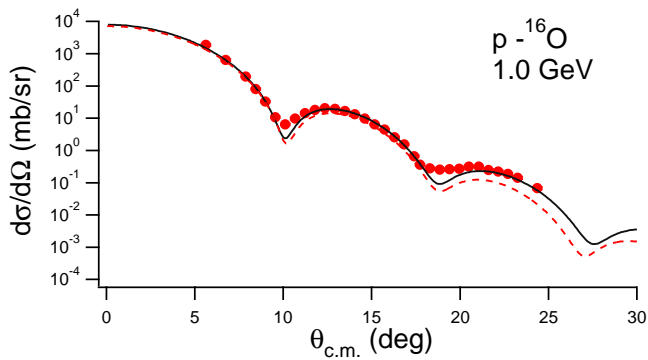}
\caption{(Color online) Comparison of the numerical results with the 
differential elastic-scattering cross section data for 
$p$\,-$^{16}$O at $E_p$=\,1.0\,GeV. 
The solid curve represents the full calculation 
with the harmonic-oscillator shell-model wave function. 
The dashed curve represents the calculation with OLA. 
The data are taken from Ref.~\cite{ozawa2001}.}
\label{16O1G}
\end{figure}

We start with the calculation of the differential 
cross section for $p$\,-$^{16}$O at an 
incident energy of 1.0 GeV, where 
the parameters of the nucleon-nucleon scattering amplitude
are well determined. 
The numerical results of the differential cross sections 
are shown in Fig.~\ref{16O1G}.
The solid curve represents the result of the full calculation 
with the harmonic-oscillator shell-model wave function, while 
the dashed curve shows the cross section calculated in the OLA.
Both results agree satisfactorily with the data. 
The two calculations give no large 
difference up to the second minimum. 
This analysis confirms that the $p$\,-$^{16}$O 
elastic-scattering cross section is reproduced 
using the wave function which has a correct size together with 
the profile function used in the present study.

\begin{figure}[t]
\epsfig{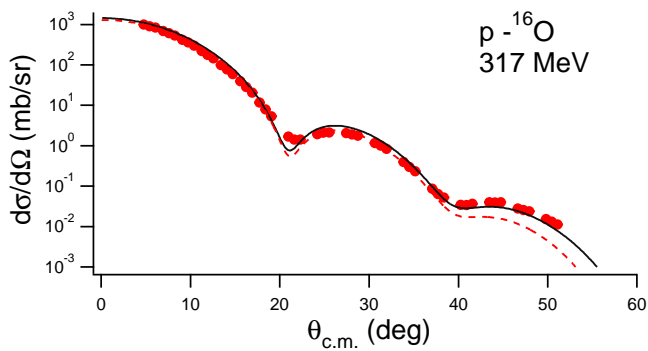}
\caption{(Color online) Comparison of the numerical results with the 
differential elastic-scattering cross section data for 
$p$\,-$^{16}$O at $E_p$=\,317\,MeV. 
See the caption of Fig.~\ref{16O1G}.}  
\label{16O317MeV}
\end{figure}

As the experimental study of $p$\,-nucleus elastic scattering at 
around 300\,MeV is considered an interesting and promising 
project at RIKEN, it is useful to assess the 
suitability of the profile function at this energy as well. 
For this purpose  
we calculate the differential cross section 
of the $p$\,-$^{16}$O elastic scattering in this energy region.  
Figure~\ref{16O317MeV} displays the results of the differential 
cross section. Both the exact and OLA curves reproduce 
the data fairly well, similarly to the case of 1 GeV.
From these comparisons at 300 MeV and 1 GeV, 
we can conclude that the $p$\,-nucleus elastic scattering 
can reliably be described in the OLA at least 
up to the second minimum 
using the present parameters of the profile function.

\subsection{The cross sections of $^{21,23}$O}

\begin{table}[b]
\caption{The rms radii, in fm,  
of the matter, neutron and proton density distributions 
for several oxygen isotopes. The  
$\nu_p$ and $\nu_n$ values given in fm are the size parameters of 
the harmonic-oscillator functions for proton and neutron. 
See Appendix~A. 
The seventh and eighth columns denote 
the total reaction cross sections, in units of mb, 
calculated in the OLA  
for proton and $^{12}$C targets at the incident energy of 
1.0\,GeV per nucleon. 
The interaction cross sections taken from Ref.~\cite{ozawa2001} 
are given in parentheses.} 
\label{radii}
\begin{center}
\begin{tabular}{cccccccl}
\hline\hline
 & $\nu_{p}$ & $\nu_{n}$ & $r_m$ & $r_n$ & $r_p$ 
& $\sigma_{\rm R}$(with $p$)  & \ \  $\sigma_{\rm R}$(with $^{12}$C) \\
\hline
$^{16}$O &  1.71 & 1.71  & 2.51  & 2.51  & 2.51 & 305 & 1003 (982$\pm$6) \\
$^{20}$O &  1.70 & 1.83  & 2.78  & 2.94  & 2.51 & 369 & 1135 (1078$\pm$10)\\
$^{21}$O &  1.70 & 1.87  & 2.85  & 3.04  & 2.51 & 386 & 1173 (1098$\pm$11)  \\   
         &  1.70 & 1.67  & 2.64  & 2.72  & 2.51 & 357 & 1098 \\ 
$^{23}$O &  1.70 & 1.98  & 3.05  & 3.30  & 2.51 & 429 & 1267 (1308$\pm$16) \\
         &  1.70 & 2.05  & 3.13  & 3.41  & 2.51 & 440 & 1302 \\
         &  1.70 & 2.15  & 3.25  & 3.58  & 2.51 & 457 & 1353 \\ 
\hline\hline
\end{tabular}
\end{center}
\end{table}

As was shown in the previous section, 
the total reaction cross sections of the oxygen isotopes 
with a $^{12}$C target are reproduced reasonably well, 
except for the cases of $^{21,23}$O. 
In particular, the total reaction cross section of 
$^{23}$O+$^{12}$C is calculated to be about 2$\%$ 
smaller than the measured interaction cross section. 
This may suggest that the $\langle S_n \rangle$ model 
slightly underestimates the matter radius of $^{23}$O.
As mentioned at the beginning of the present section, 
we increase the radius to reproduce the interaction cross 
section. Three sets of $\nu_p$ and $\nu_n$ values are 
listed in Table~\ref{radii}. In the case of $^{21}$O, 
the total reaction cross section calculated with the 
NTG approximation is about 5\% 
larger than the interaction cross section. As $^{21}$O has 
four particle-bound excited states, the difference between the 
two cross sections could be reduced to some extent. 
Table~\ref{radii} lists a set which fits the interaction cross 
section of $^{21}$O+$^{12}$C.    

\begin{figure}[t]
\epsfig{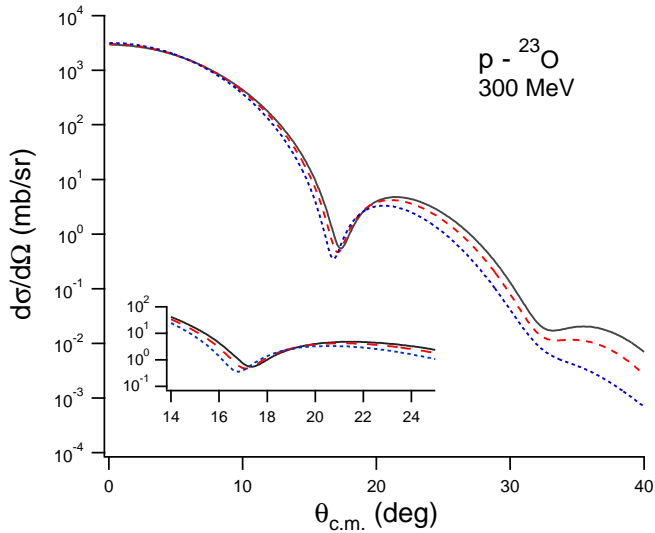}
\caption{(Color online) Comparison of the differential 
elastic-scattering cross sections of 
$p$\,-$^{23}$O at $E_{p}$=\,300\,MeV for different values of 
nuclear matter radii. 
The solid curve represents the case with a radius of 3.05 fm, 
the dashed curve with a radius of 3.15 fm, 
and the dotted curve with a radius of 3.25 fm. 
The calculations are carried out with the OLA.}
\label{23O300MeV}
\end{figure}

Now we examine whether or not the $p$\,-nucleus 
elastic scattering can give us useful information on the 
matter radius of the nucleus. Figure~\ref{23O300MeV}
displays the differential cross 
section of $p$\,-$^{23}$O elastic-scattering 
at $E_p$=300\,MeV for the three different matter radii listed in 
Table~\ref{radii}. We focus on the angle of the first peak 
and the magnitude of the cross section at this angle. 
As the matter radius increases, the first peak moves to a smaller 
angle. The angle in degrees (the magnitude of the cross section in 
mb/sr) of the first peak for the solid, dashed, and dotted 
curves are 21.5 (4.7), 21.1 (4.2), and 20.5 degree (3.3 mb/sr), 
respectively.

The difference, at each first peak angle, 
between the magnitudes of the cross sections denoted 
by the solid and dotted curves is about 30$\%$. 
This value is not small, but, unfortunately  
the presently estimated experimental uncertainty at this angle 
for the $p$\,-$^{20}$O elastic scattering is 
comparable to this value~\cite{tera}.
The angle shift of the first peak position between the two curves 
is about 1.0 degree, which might be too small 
to distinguish experimentally. 
 
\begin{figure}[b]
\epsfig{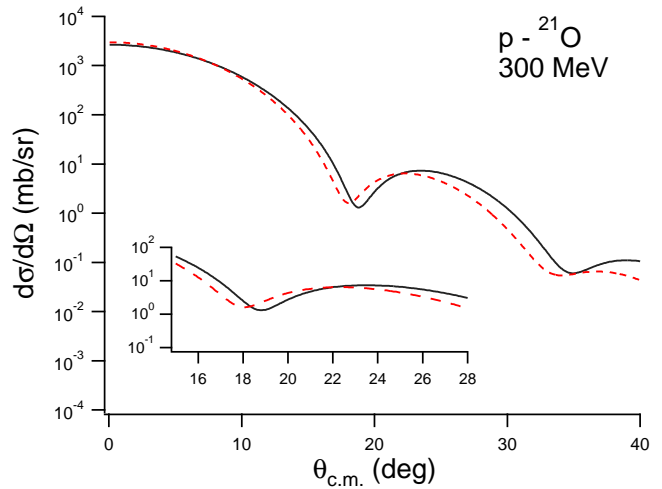}
\caption{(Color online) Comparison of the differential 
elastic-scattering cross sections of 
$p$\,-$^{21}$O at $E_{p}$=\,300\,MeV for different values of 
nuclear matter radii. 
The solid curve represents the case with a radius of 2.64 fm, 
while the dashed curve with a radius of 2.85 fm. 
The calculations are carried out with the OLA.}
\label{21O300MeV}
\end{figure}

We have carried out a similar analysis for the case of 
$^{21}$O. The result is displayed in Fig.~\ref{21O300MeV}. 
The angle (the magnitude of the cross section) of the first peak 
for the solid and dashed curves are 
23.5 (7.3), 22.3 degree (6.4 mb/sr), respectively.
Again a high precision experiment will be needed to distinguish 
which of the above two is favorable.

\subsection{Prediction of the cross section of $^{20}$O}

Recently the differential cross section of 
$^{20}$O elastic scattering on a proton 
has been measured at the energy of 300\,MeV per nucleon 
at NIRS~\cite{tera}. 
Motivated by this experiment, we predict the 
elastic and total reaction cross sections of $p$\,-$^{20}$O.  

\begin{figure}[t]
\epsfig{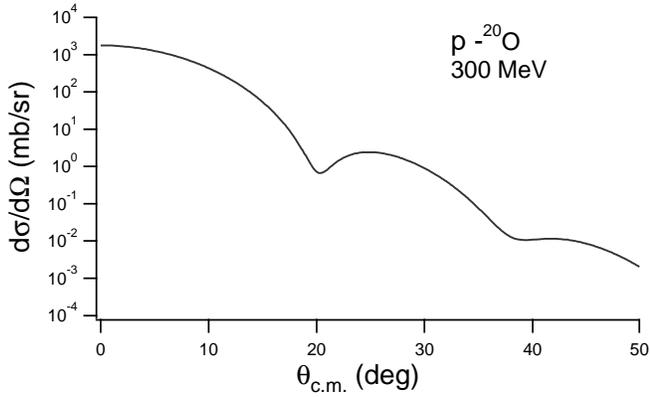}
\caption{Prediction of the differential elastic-scattering 
cross section 
for $p$\,-$^{20}$O at $E_p$=\,300\,MeV. 
The curve represents the full calculation 
with the harmonic-oscillator shell-model wave function.}
\label{20O300MeV}
\end{figure}

The Glauber amplitude is calculated fully 
using the Slater determinant consisting of the 
harmonic-oscillator single-particle orbits. 
The size parameter of the harmonic oscillator wave function 
is chosen to reproduce the radius of $^{20}$O given in the 
$\langle S_n \rangle$ model, as listed in Table~\ref{radii}. 
The result is shown in Fig.~\ref{20O300MeV}. The first peak 
appears at 24.8\,$^{\circ}$ and the cross section at the 
peak is 2.43 mb/sr. As shown in Fig.~\ref{reac.oxygen}, 
the  reaction cross section calculated in the NTG approximation 
is about 30\,mb 
larger than the interaction cross section. It is not clear, 
however, whether this difference between the two 
cross sections indicates that 
the radius used for $^{20}$O is slightly too large or not, 
because a considerable contribution to the reaction cross section 
is expected from the 12 particle-bound excited states of $^{20}$O. 
In any case the predicted differential cross section 
near the peak appears 
consistent with the preliminary data~\cite{tera}.

\section{Summary}
\label{summary}

We have systematically calculated 
the total reaction cross sections 
of the oxygen isotopes, $^{15-24}$O, on a $^{12}$C target 
at high energies using the Glauber theory. 
The oxygen isotopes are described with the Slater determinants 
generated from a phenomenological mean-field potential, 
which is an extension of 
our previous work for describing the carbon 
isotopes~\cite{hsbk,bhks}. 
We have introduced two schemes, 
the $S_n$ model and the $\langle S_n \rangle$ model. 
The agreement between theory and experiment is generally good, 
especially in the $\langle S_n \rangle$ model,  
but the sharp increase of the reaction cross sections 
from $^{21}$O to $^{23}$O remains an open question. 

As a possible cross section which may depend on the 
nuclear radius more sensitively than the reaction cross 
section, we have examined the differential 
cross sections of the $p$-$^{20,21,23}$O elastic-scatterings 
at the incident energy of $E_p =$ 300\,MeV 
using the full and OLA Glauber amplitudes. The 
differential cross sections 
calculated from the two amplitudes are not very different 
up to the second minimum of the angular distribution. 
We have calculated the total reaction cross sections 
by varying the nuclear radii so as 
to be consistent with the observed interaction cross sections, 
and then analyzed the sensitivity of the differential cross 
sections to the radii. 
We find a considerable change in the cross section, 
but whether it can really be observed or not 
strongly depends on the precision of the experimental data. 

We have predicted the differential  
cross section for $p$\,-$^{20}$O elastic-scattering, which has 
recently been measured.  
Our prediction appears consistent 
with the preliminary data. 
This implies that it is possible to calculate 
to a good approximation 
the total reaction and
elastic scattering cross sections at high energies.

\vspace{5mm}

We are grateful to B. V. Carlson for his careful reading of the 
manuscript. We 
acknowledge T. Motobayashi for his encouragement
during the course of this work. 
B. A-I. was supported in part by Japan International Cultural 
Exchange Foundation (JICEF). 
W. H. is supported by the Japan Society 
for the Promotion of Science for Young Scientists.
This work was in part supported by a Grant for Promotion of 
Niigata University Research Projects (2005--2007), 
and a Grant-in Aid for 
Scientific Research for Young Scientists (No. 19$\cdot$3978). 
One of the authors (Y. S.) thanks the JSPS core-to-core program, 
Exotic Femto Systems.

\appendix
\section{Harmonic-Oscillator Density}

We briefly explain the method of calculating the neutron or proton 
intrinsic density which is employed in our paper~\cite{hsbk}. 
We assume a Slater determinant constructed from 
harmonic-oscillator wave functions.  
Denoting the Slater determinant by $\Psi$, we first calculate 
the neutron or proton density which contains the effect of 
the center-of-mass motion
\begin{equation}
\tilde{\rho}({\vi r})=\langle\Psi\vert
\sum_{i=1}^{A}\delta({\vi r}_{i}-{\vi r})P_{i}\vert\Psi\rangle,
\end{equation}
where ${\vi r}_{i}$ is the nucleon single-particle coordinate and 
$P_{i}$ is a projector to either neutron or proton.  
The single-particle wave function is given 
by the harmonic-oscillator 
function
\begin{equation}
\psi_{nljm}=R_{nlj}(r)[Y_{l}(\hat{\vi r})\chi_{1/2}]_{jm},
\end{equation}
where $\chi_{1/2}$ denotes the spin function. If the orbit specified 
by $nlj$ is not completely filled, we take an average over the $z$ 
component $m$, obtaining a spherical density. 
After some angular momentum algebra we have 
\begin{equation}
\tilde{\rho}({\vi r})=\frac{1}{4\pi}
\sum_{nlj}{N}_{nlj}R_{nlj}^{2}(r),
\end{equation}
where $N_{nlj}$ is the occupation number of the $nlj$ orbit. 

To obtain the intrinsic density which has no contribution from the 
center-of-mass motion, we make use of the fact that 
the center-of-mass  
motion $\Psi_{\rm cm}({\vi X})$ contained in $\Psi$ is factored 
out as
\begin{equation}
\Psi=\Psi_0\Psi_{\rm cm}({\vi X}),
\end{equation}
where ${\vi X}$ is the center-of-mass coordinate and 
\begin{equation}
\Psi_{\rm cm}({\vi X})=\left(\frac{A}{\pi{\nu}^2}\right)^{3/4}
\exp\Big(\!-\!\frac{A}{2{\nu}^2}{\vi X}^2\Big).
\end{equation}
Then the intrinsic density ${\rho}({\vi r})$ defined by 
\begin{equation}
{\rho}({\vi r})=\langle\Psi_0\vert
\sum_{i=1}^{A}\delta({\vi r}_{i}-{\vi X}-{\vi r})P_{i}
\vert\Psi_0\rangle
\end{equation}
is obtained from the following relation~\cite{hsbk}
\begin{equation}
\int d{\vi r}\,{\rm e}^{i{\vis
 k}\cdot{\vis r}}\rho({\vi r})=[\langle \Psi_{\rm cm}|{\rm e}^{i{\vis
 k}\cdot{\vis X}}| \Psi_{\rm cm}\rangle]^{-1}
\int d{\vi r}\,{\rm e}^{i{\vis k}\cdot{\vis r}}\tilde{\rho}({\vi r}).
\label{int.density}
\end{equation}

The radial functions which we need for the oxygen isotopes are 
\begin{eqnarray}
R_{00}(r)&=&\Big(\frac{4}{\sqrt{\pi}\nu^{3}}\Big)^{1/2} 
{\rm exp}\Big(\!-\!\frac{r^{2}}{2\nu^{2}}\Big),
\nonumber\\
R_{01}(r)&=&\Big(\frac{8}{3\sqrt{\pi}\nu^{5}}\Big)^{1/2} r\, 
{\rm exp}\Big(\!-\!\frac{r^{2}}{2\nu^{2}}\Big),
\nonumber\\
R_{02}(r)&=&\Big(\frac{16}{15\sqrt{\pi}\nu^{7}}\Big)^{1/2} r^{2}\,
{\rm exp}\Big(\!-\!\frac{r^{2}}{2\nu^{2}}\Big),\nonumber\\
R_{10}(r)&=&\Big(\frac{8}{3\sqrt{\pi}\nu^{3}}\Big)^{1/2}
\Big(\frac{r^{2}}{\nu^{2}}-\frac{3}{2}\Big)\,
{\rm exp}\Big(\!-\!\frac{r^{2}}{2\nu^{2}}\Big).
\end{eqnarray}
We give the examples of $^{21,23}$O below. 
The neutron or proton density of $^{21}$O is 
\begin{eqnarray}
\tilde{\rho}({\vi r})
&\!\!=\!\!&\frac{1}{4\pi}\Big[2R^{2}_{00}+6R^{2}_{01}+(M-8)R^{2}_{02}\Big]
\nonumber\\
&\!\!=\!\!&\frac{2}{\pi\sqrt{\pi}\nu^{3}}\,
{\rm exp}\Big(\!-\!\frac{r^{2}}{\nu^{2}}\Big)\Big[1+2\frac{r^{2}}{\nu^{2}}+
\frac{2(M-8)}{15}\frac{r^{4}}{\nu^{4}}\Big],\nonumber \\
\end{eqnarray}
were $M$ is the number of protons $Z$=8 or the number of neutrons $N$=13. 
The neutron density for $^{23}$O is 
\begin{eqnarray}
\tilde{\rho}({\vi r})
&\!\!=\!\!&\frac{1}{4\pi}\Big[2R^{2}_{00}+6R^{2}_{01}+6R^{2}_{02}+R^{2}_{10}\Big]
\nonumber\\
&\!\!=\!\!&\frac{2}{\pi\sqrt{\pi}\nu^{3}}\,
{\rm exp}\Big(\!-\!\frac{r^{2}}{\nu^{2}}\Big)
\Big[\frac{7}{4}+\frac{r^{2}}{\nu^{2}}+
\frac{17}{15}\frac{r^{4}}{\nu^{4}}\Big].
\end{eqnarray}

The corresponding intrinsic density is obtained from the relation 
(\ref{int.density}). The results are 
\begin{eqnarray}
\rho({\vi r})&=&{2}\Big(\frac{u}{\pi \nu^{2}}\Big)^{3/2}
{\rm exp}\Big(\!-\!\frac{ur^{2}}{\nu^{2}}\Big)\nonumber\\
&\ \ \times& \!\!\Big[1+3(1-{u}) +\frac{15}{2}\alpha(1-u)^2\nonumber\\
&&+{2u^2}(1\!+\!5\alpha(1\!-\!u))\frac{r^{2}}{\nu^{2}}
+2\alpha u^4\frac{r^{4}}{\nu^{4}}\Big],
\label{dens.21O}
\end{eqnarray}
for $^{21}$O and 
\begin{eqnarray}
\rho({\vi r})&=&{2}\Big(\frac{u}{\pi\nu^{2}}\Big)^{3/2}
{\rm exp}\Big(\!-\!\frac{ur^{2}}{\nu^{2}}\Big)
\Big[
\frac{15}{2}-{10u}+\frac{17u^2}{4}\nonumber\\
&&+\Big(\frac{20u^2}{3}-\frac{17u^3}{3}\Big)\frac{r^{2}}{\nu^{2}}
+\frac{17u^4}{15}\frac{r^{4}}{\nu^{4}}\Big],
\label{dens.23O}
\end{eqnarray}
for $^{23}$O.  Here $u=A/(A-1)$ and $\alpha=(M-8)/{15}$.

As is evident, the above derivation assumes that 
the harmonic-oscillator
constant $\nu$ is the same for protons and neutrons. When we use 
different values for these, as in Table~\ref{radii}, 
the intrinsic densities of 
Eqs.~(\ref{dens.21O}) and (\ref{dens.23O}) 
will no longer be exact. However, we have still used 
these equations to calculate the intrinsic densities in that case.

\section{Elastic scattering amplitude}

In this appendix, following the method of Refs.~\cite{bassel}
we explain our calculation scheme 
of the proton-nucleus elastic scattering amplitude 
for a Slater determinant wave function 
constructed from harmonic-oscillator single-particle orbits. 

Let us start with the $p$\,-nucleus elastic-scattering 
amplitude already given by Eq.~(\ref{glamp}), 
\begin{equation}
   F({\vi q})=\frac{iK}{2 \pi}\int 
   d{\vi b}\,{\rm e}^{i{\vis {q\cdot b}}}
   \left(1-{\rm e}^{i\chi_N({\vis b})}\right),
\label{glampa}
\end{equation}
where ${\vi b}$ is the impact parameter vector
perpendicular to the beam ($z$) direction, 
${\vi K}$ the initial momentum of the relative motion,  
and ${\vi q}$ the momentum 
transferred from the projectile to the target.

As we have shown in Eq.~(\ref{opsf}), $\chi_N({\vis b})$ is 
the nuclear part of the optical phase-shift function 
given by 
\begin{eqnarray}
   {\rm e}^{i\chi_N({\vi b})}
   &\!\!=\!\!& \langle\Psi_{0}\vert
     \prod_{i=1}^{A}
     \Big[1 - \frac{1+\tau_{3_i}}{2}
              \Gamma_{pn}({\vi b} - {\vi s}_{i})
\nonumber \\
     & & \ \ \qquad \qquad 
          -   \frac{1-\tau_{3_i}}{2}
              \Gamma_{pp}({\vi b} - {\vi s}_{i}) 
     \Big]\vert \Psi_{0} \rangle 
\nonumber\\
   &\!\!=\!\!& \langle\Psi_{0}\vert
     O({\vi b} - {\vi s})\vert \Psi_{0} \rangle, 
\label{oper}
\end{eqnarray}
where $\Psi_{0}$ is the intrinsic (translation-invariant) 
$A$-nucleon wave function of the nuclear ground state. 
The coordinate 
${\vi s}_{i}$ is the projection onto the $xy$-plane 
of the nucleon coordinate relative to 
the center-of-mass of the nucleus, {\it i.e.}, 
${\vi s}_i={\vi s}_i^{\prime}-{\vi X}$, where 
${\vi s}_i^{\prime}$ is the coordinate of a nucleon 
in the nucleus projected onto the $xy$-plane, and 
${\vi X}$ is the center-of-mass coordinate of the nucleus. 
$\tau_{3_i}$ is $1$ for neutron and $-1$ for proton.
For later convenience, here we introduce an operator 
$O({\vi b} - {\vi s})$.

The main idea is that, for the harmonic-oscillator
Slater determinant wave function, 
the elastic scattering amplitude $F^{\prime}({\vi q})$ 
referring to the coordinate origin is factorized 
into a center-of-mass part and the intrinsic
amplitude $F({\vi q})$~\cite{bassel}, 
where $F^{\prime}({\vi q})$ is defined by 
\begin{eqnarray}
F^{\prime}({\vi q})&=&\frac{iK}{2 \pi}\int
d{\vi b}\,{\rm e}^{i{\vis q}\cdot{\vis b}}
\langle\Psi\vert \delta^{(3)}({\vi X} - (1/A)\sum_j {\vi r}_j)
\nonumber\\ 
 &{}&  \times \left(1- O({\vi b} - {\vi s}^{\prime}) \right)
\vert\Psi\rangle,  
\label{sep}
\end{eqnarray}
For the harmonic oscillator, 
$\vert\Psi\rangle$ is exactly factorized as 
$\vert\Psi\rangle$ $= \vert\Psi_{cm}\rangle \vert\Psi_0\rangle$.

Then, the intrinsic amplitude which is free from the center-of-mass 
motion is easily calculated from $F^{\prime}({\vi q})$ as
\begin{eqnarray}
   F^{\prime}({\vi q}) &=& \langle\Psi_{\rm cm}\vert
                           e^{-i{\vis q}\cdot {\vis X}}
                           \vert\Psi_{\rm cm}\rangle F({\vi q}),
\label{sep2}
\end{eqnarray}
where
\begin{equation}
 \langle\Psi_{\rm cm}\vert
e^{-i{\vis q}\cdot {\vis X}}
\vert\Psi_{\rm cm}\rangle
={\rm exp}[-\nu^{2}q^{2}/4A], 
\end{equation}
and $\nu$ is the size parameter of the harmonic-oscillator 
potential.  
The separability of the elastic scattering amplitude, 
Eq.~(\ref{sep}), is valid when the size parameter of 
the harmonic well for protons, $\nu_{p}$, is
equal to that for the neutrons, $\nu_{n}$, 
which is the case for $^{16}$O. 
On the other hand, when $\nu_{p}$ $\neq \nu_{n}$,
which is the case for $^{20,21,22}$O, 
we only assume this separability and adopt 
the following relation \cite{hsbk}:  
\begin{equation}
\nu^{2} = \frac{Z}{A} \nu_{p}^{2} + \frac{N}{A} \nu_{n}^{2}.
\end{equation}

Now, we focus on the calculation of 
the elastic scattering amplitude $F^{\prime}({\vi q})$ 
using the nucleon coordinates referring to the coordinate origin. 
Since we represent the ground-state wave function 
as a Slater determinant, 
\begin{equation}
\Psi = (A!)^{-1/2}{\rm det}[\psi_{m}(r_{n}^{\prime})], 
\end{equation}
the antisymmetrization is only required for the bra, such as 
\begin{equation}
\langle\Psi\vert=(A!)^{-1/2}
\rm det [\delta_{mn}\psi_{m}^{\dag}(r_{n}^{\prime})].
\end{equation}

As one can see from Eq.~(\ref{oper}), 
the operator $O({\vi b} - {\vi s}^{\prime})$
is factorizable into the operators 
that act on one particle subspaces, so that
\begin{equation}
\langle\Psi\vert
O({\vi b}-{\vi s}^{\prime})
\vert \Psi \rangle
={\rm det} [O_{nm}]_{n} {\rm det} [O_{nm}]_{p},
\label{opsf-1}
\end{equation}
where ${\rm det}[O_{nm}]_{p}$ is the $Z\times Z$
determinant for $Z$ protons (in this work, $Z=8$)
and 
${\rm det}[O_{nm}]_{n}$ is the $N\times N$
determinant for $N$ neutrons. 
$O_{nm}$ is the matrix element defined by
\begin{equation}
O_{nm}=\delta_{nm}-\int{\psi^{\ast}_{m}({\vi r}^{\prime})}
\Gamma({\vi b}+{\vi s}^{\prime})
\psi_{n}({\vi r}^{\prime})d{\vi r}^{\prime}.
\end{equation}

We have calculated the matrix element, 
$O_{nm}$, analytically and
obtained the determinants~\cite{bassel}. 
Concerning the neutron part, ${\rm det}[O_{nm}]_{n}$, 
we take the average of 
the determinants of different configurations. 
For example, for $^{20}$O, which has a $0^{+}$ ground state,
we have 12 neutrons, four of which are in the $0d_{5/2}$ state. 
Therefore, we have three different determinants 
that give a $0^{+}$ ground state of $^{20}$O.

\end{document}